\documentclass[preprint,showpacs,amsmath,amssymb]{revtex4}

\usepackage{graphicx,psfrag}
\usepackage{amssymb}

\begin{document}

\title{Search for optimal 2D and 3D wave launching configurations for the largest acceleration
of charged particles in a magnetized plasma, Resonant Moments
Method}

\author{M. Ponomarjov}
\author{D. Carati}
\affiliation{Statistical and Plasma Physics, CP 231, Campus
Plaine, Universit\'e Libre de Bruxelles, 1050 Bruxelles, Belgium.}

\date{\today}

\begin{abstract}
Optimal two-dimensional (2D), three-dimensional (3D) wave
launching configurations are proposed for enhanced acceleration of
charged particles in magnetized plasmas. A primary wave is
launched obliquely with respect to the magnetic field and a
secondary, low amplitude, wave is launched perpendicularly. The
effect of both the launching angle of the primary wave, and the
presence of the secondary wave is investigated. Theoretical
predictions of the highest performances of the three-dimensional (3D)
configurations are proposed using a Resonance Moments Method (RMM)
based on estimates for the moments of the velocity distribution
function calculated inside the resonance layers (RL). They suggest
the existence of an optimal angle corresponding to non parallel
launching. Direct statistical simulations show that it is possible
to rise the mean electron velocity up to the order of magnitude as
compared to the primary wave launching alone. It is a quite
promising result because the amplitude of the secondary wave is
ten times lower the one of the first wave. The parameters used are
related to magnetic plasma fusion experiments in electron
cyclotron resonance heating and electron acceleration in planetary
ionospheres and magnetospheres.
\end{abstract}


\maketitle Electron acceleration due to external radio frequency
waves in a strong magnetic field has long been recognized as an
important effect in a wide variety of problems ranging from plasma
heating and current drive in fusion
devices~\cite{fis87,cai91,erga94,llo98} to electron acceleration
in the Earth's radiation belts during geomagnetic storms
~\cite{Horne, Summers}, active ionospheric and magnetospheric
probing ~\cite{bamage95,chdadr01}.

It is well known that wave-particle interactions are most
efficient when the particles are in resonance with the waves. The
resonance conditions
\begin{equation}
\label{resonance}
k_\parallel v_\parallel - \omega = \frac{N\Omega}{\gamma}
%
\end{equation}
define some regions 'of sensitivity' in wave-particle parameter
space which can be described as Resonant Layers (RL).
Here,
$N$ is the harmonic number, $\omega=2\pi f$ is the wave frequency,
$k_\parallel$ and $v_\parallel$ are the components of the wave
vector and the electron velocity parallel to the constant magnetic
field $B_0$,  $\gamma=(1-v^2/c^2)^{-1/2}$ is the relativistic
factor and $\Omega=e B_0/m_e$ the gyro-frequency. This view allows
to develop significantly previous simulations of charged particle
fluxes and plasma disturbances in ambient magnetic field in
\cite{Ponomarjov02,Ponomarjov01,Ponomarjov00,
Ponomarjov97,Ponomarjov96}.
 There is a special case when electrons
permanently staying in the RL~(\ref{resonance}). Such a phenomenon
has been referred to as autoresonance~\cite{dav62,kole63,robu64}
and its conditions are known as the cyclotron auto-resonance maser
(CARM) conditions~\cite{cocomc91}. Several mechanisms have been
explored for maintaining the synchronization between electrons and
waves not fully satisfying these CARM conditions such as changing
the profile of the guide magnetic field or varying the wave phase
velocity~\cite{fri94,mil97}. Recently, the use of two parallel
counter-propagating waves has been
considered~\cite{gena98,gena99}. Numerical tests~\cite{locage01}
have shown that the two-wave scheme may lead to higher averaged
parallel velocity. The stochastic acceleration mechanism for
electrons in a plane monochromatic electromagnetic wave
propagating obliquely to the external magnetic field has also been
studied~\cite{medrpa87,vibu87,kaakom90}. It was found that it is
easier to accelerate electrons to high energies with increasing
propagation angle when the electron motion becomes stochastic and
the parallel phase velocity of wave is supraluminous
($n_{\parallel}=k_{\parallel }c/\omega < 1$). Furthermore,
Karimabadi and Angelopoulos÷\cite{kaan89} studied the interaction
of charged particles with a packet of obliquely propagating plane
monochromatic electromagnetic waves under the special condition
(of equal $n_i\cos \theta_i$ for all the waves, where for the
$i$-wave $n_i$ - refraction index, $\theta_i$ - the propagation
angle). This condition allowed the system to be reduced to two
degrees of freedom and the particles can be accelerated through a
process of Arnol'd diffusion~\cite{arn64} in two dimensions.

The majority of the existing works are based on the description of
a single particle dynamics in one (or more but under condition of
equal $n_i\cos \theta_i$) plane monochromatic radio frequency
waves.

The coherent acceleration of nonrelativistic ions interacting with
two and more electrostatic waves in a uniform magnetic field has
been studied by \cite{Ram}, recently by \cite{Strozzi}, as a
generalization of analysis of \cite{Benisti1,Benisti2} by
including wavenumbers along the external magnetic field.

\begin{figure}[h]
\includegraphics[width=0.65\columnwidth]{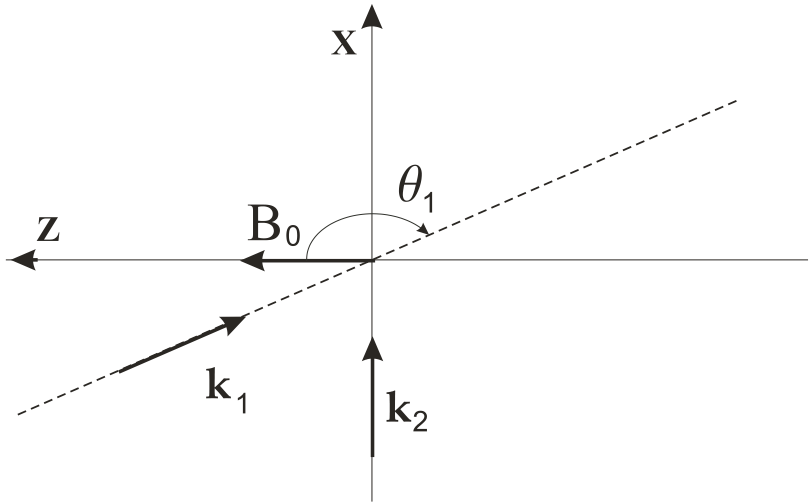}
\caption{Schematic picture of the electro-magnetic configuration.}
\label{config}
\end{figure}

In this paper, a mechanism is discussed for the acceleration of
electron populations resulting from the effect of crossing
electromagnetic waves propagating in a dispersive medium according
to the geometry represented in Figure~\ref{config} (the condition
of equal $n_i\cos \theta_i$ for the two waves is thus clearly
broken). To analyze this mechanism, the resonance moments (RM) of
the distribution, i.e. velocity moments computed in the RL only,
are evaluated. The first order RM suggests that a peculiar
$\theta$ results in a maximal averaged parallel flux. Although the
RM approach has to be considered as an approximation, this
prediction is reasonably confirmed by direct statistical
simulations. Moreover, the two-wave scheme allows to rise the mean
electron velocity up to one order of magnitude when compared to
the one-wave scheme, based on the primary wave only.

The electromagnetic configuration that is considered
(Figure~\ref{config}) is the combination of a strong magnetic
field (assumed to be along the $z$ direction), a primary wave
propagating obliquely with respect to the magnetic field and a
secondary wave propagating perpendicularly to the magnetic field.
As a first step, to simplify direct particle simulations both the
primary and the secondary waves are assumed to be in the plane
$(x,z)$. This simplification can give not so impressive effect as
compared to 3D launching. Nevertheless, it will give a first
estimate of the secondary wave effect and motivation to develop
more realistic 3D launching code that will be closer to real
experimental setups.

This electromagnetic configuration is not an attempt to satisfy the resonance condition during a
long time being close to the cyclotron auto-resonance maser (CARM)
conditions~\cite{cocomc91}. Rather, a large population of
electrons is considered and only the average effect of the waves
on the population is considered. The fraction of electrons that
are close to the condition~(\ref{resonance}), that corresponds to
a resonance layer (RL) in the velocity space, becomes then as
important as the time these electrons remain resonant. The
secondary wave does not carry any parallel momentum and cannot
induce any net parallel motion of the electrons. The purpose of
this secondary wave is to maintain a pseudo-equilibrium velocity
distribution in which the RL is continuously re-filled. Indeed,
the combined effect of the two-waves and the magnetic field yields
a stochastic motion during which the synchronization between the
waves and the gyro-motion of the electron is, on average, more
favorable for transferring momentum to the electrons.

\begin{figure}[h]
\includegraphics[width=0.65\columnwidth]{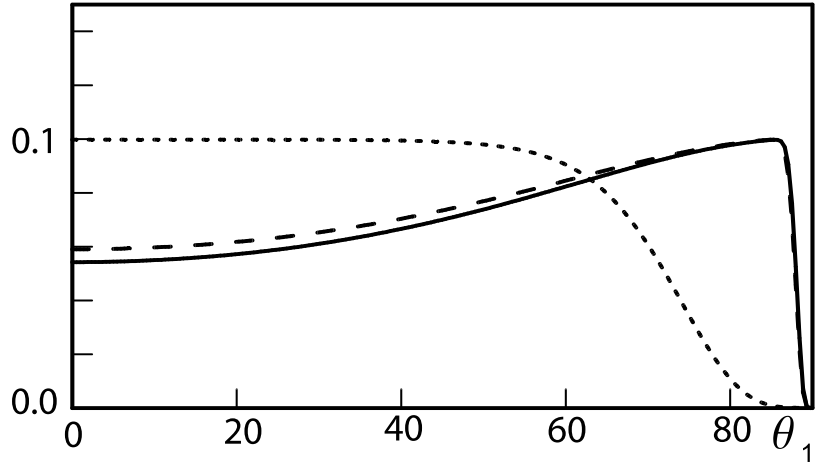}
\caption{Averaged parallel flux of particles in the RL in arbitrary units for a Maxwellian distribution for increasing densities, $e_0=0.3$ solid line,  $e_0=0.6$ dashed line  and $e_0=1.99$ dotted line.}
\label{averaged-flux}
\end{figure}

  The description of the electron trajectory $(\mathbf{r},
\mathbf{p})$ requires a relativistic treatment and is derived from the time dependent
Hamiltonian:
\begin{equation}
\label{Ham}
H=\sqrt{m^{2}c^{4}+c^2(\mathbf{p}+e\mathbf{A})^{2}}\,,
\end{equation}
and the trajectories of the electrons are determined by the initial conditions and by the
Hamilton equations.
This picture corresponds to a test-electron in an external electromagnetic field.
Assuming that the electrons interact with two monochromatic waves propagating in cold plasma
at angles $\theta_1$ and $\theta_2$ with respect to the guide magnetic field, the total vector
potential can be written as follows:
\begin{eqnarray}
\mathbf{A}&=&B_0\ x\ \mathbf{e}_y\nonumber+\frac{A_1}{2}\ e^{i(\mathbf{k}_1.\mathbf{r}-\omega_1 t)}\ \left(\cos\psi_1 \mathbf{e}_y-i\sin\psi_1 \mathbf{e}_1\right)  \nonumber \\
&+&\frac{A_2}{2}\ e^{i(\mathbf{k}_2.\mathbf{r}-\omega_2 t)}\ \left(\cos\psi_2 \mathbf{e}_y-i\sin\psi_2 \mathbf{e}_2\right) +c.c.\, ,
\label{vecpot}
\end{eqnarray}
where $\mathbf{e}_1$ and $\mathbf{e}_2$ are two unit vectors in the plane $(x-z)$.
 These vectors, as well as the angles $\psi_1$ and
$\psi_2$ and the refraction indices $n_1=\|\mathbf{k}_1\|
c/\omega_1$ and $n_2=\|\mathbf{k}_2\| c/\omega_2$, are determined
by the Appleton-Hartree dispersion relation in the cold
homogeneous plasma approximation~\cite{sti92}. Collisions with
other particles, electrons or ions, are neglected.  The
dynamics of the electrons in the electromagnetic
configuration~(\ref{vecpot}) is chaotic and unpredictable
analytically in general. Electrons with slightly different initial
position and velocity may experience drastically different
evolutions. The exact analytical prediction of the average effect
of the waves is thus out of reach. It is however possible to
anticipate the existence of a RL~(\ref{resonance}) in the velocity
space to estimate the possible net effect of the waves by
computing the RM defined by:
\begin{equation}
I^g=\int_{\mathbf{v}\in\text{ res. layer}} d\mathbf{v}\ f(\mathbf{v})\ g(\mathbf{v})
\end{equation}
for any function $g$ of the velocity. It represents the density of
electrons close to the RL for $g=1$ and the mean parallel flux of
the electrons on the layer for $g=v_\parallel$. These quantities
should give some estimate on the efficiency of the electron-wave
interaction. The RL corresponds to an ellipse in the velocity
space and the integral $I$ can be evaluated analytically using
elliptical coordinates.  For instance, assuming a Maxwellian
distribution, $f(\mathbf{v})=C\exp \left( -\beta
\mathbf{v}^2\right)$, the evaluation of $I$ for $g=v_\parallel$,
which will be denoted $I^*$ can be done explicitly. This quantity
corresponds to the averaged parallel flux of the particles that
belongs to the RL. It is, at least indirectly, related to the net
averaged parallel current produced by the electron-wave
interactions. Indeed, these interactions tend to remove electrons
from the RL while the pseudo thermalization of the electrons due
to the combined effect of the two waves is to refill constantly
the layer. The thermalization is thus expected to add a net
averaged parallel velocity proportional to $I^*$. Of course, $I^*$
gives only a rough indication of the efficiency of electron-wave
interactions and the final averaged velocity cannot be deduced
directly from it. The exact expression for $I^*$ is quite long. It
is thus more illustrative to present the
Figure~\ref{averaged-flux} in which $I^*$ is a function of
$\theta_1$, the angle of the primary wave. The dimensionless wave
amplitudes and the quadratic plasma frequency are defined by
$\overline{A}_{1,2}=A_{1,2}\Omega/cB_0$,
$e_0=(\omega_{pe}/\Omega)^2$. The other parameters correspond to
the values used in the simulations described below.

\begin{figure}[h]
\begin{minipage}{\columnwidth}
\includegraphics[width=0.65\columnwidth]{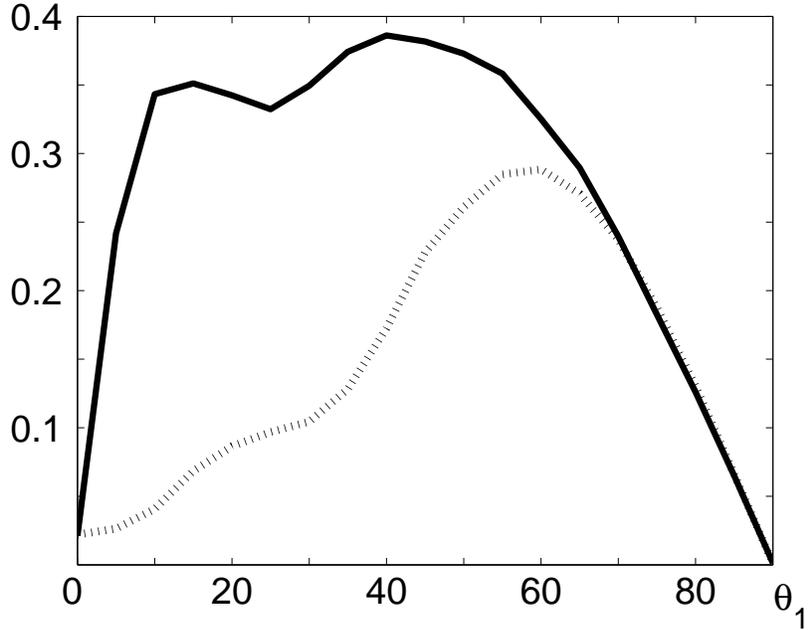}
\end{minipage}
\caption{Averaged electron velocity versus $\theta_1$ (in
degrees) for the one-wave scheme with $\overline{A}_1=0.1$,
$\overline{A}_2=0$ (dotted lines) and the two-wave-scheme with
$\overline{A}_1=0.1$, $\overline{A}_2=0.01$ (solid lines) at
time $\Omega\ t=7000$, for low-density
runs $e_0=0.3$.} \label{current-theta}
\end{figure}

As expected, for perpendicular propagation $\theta_1=90^\circ$, no
averaged parallel flux is observed. For two sets of parameters,
there is a clear maximum of the averaged parallel flux of the
particles in the RL for $\theta_1\not=0^\circ$. The explanation
for such a phenomenon is that the averaged parallel velocity
induced by the electron-wave interaction depends on both the angle
of propagation of the primary wave and the number of electrons
that are close to the RL. Smaller angles of propagation correspond
to higher parallel momentum carried on by the wave. However, the
RL condition is compatible with larger numbers of electrons for
higher angles of propagation (at least assuming a Maxwellian
distribution). It should be noted that here $I^*$ has been
computed using only the fundamental $N=1$ RL. Contributions from
the higher harmonic layers decrease rapidly because these layers
are more and more symmetric and because the absolute value of the
resonant velocity increases towards high values for which the
electron density is very small.

\begin{figure}[h]
\begin{minipage}{\columnwidth}
\includegraphics[width=0.65\columnwidth]{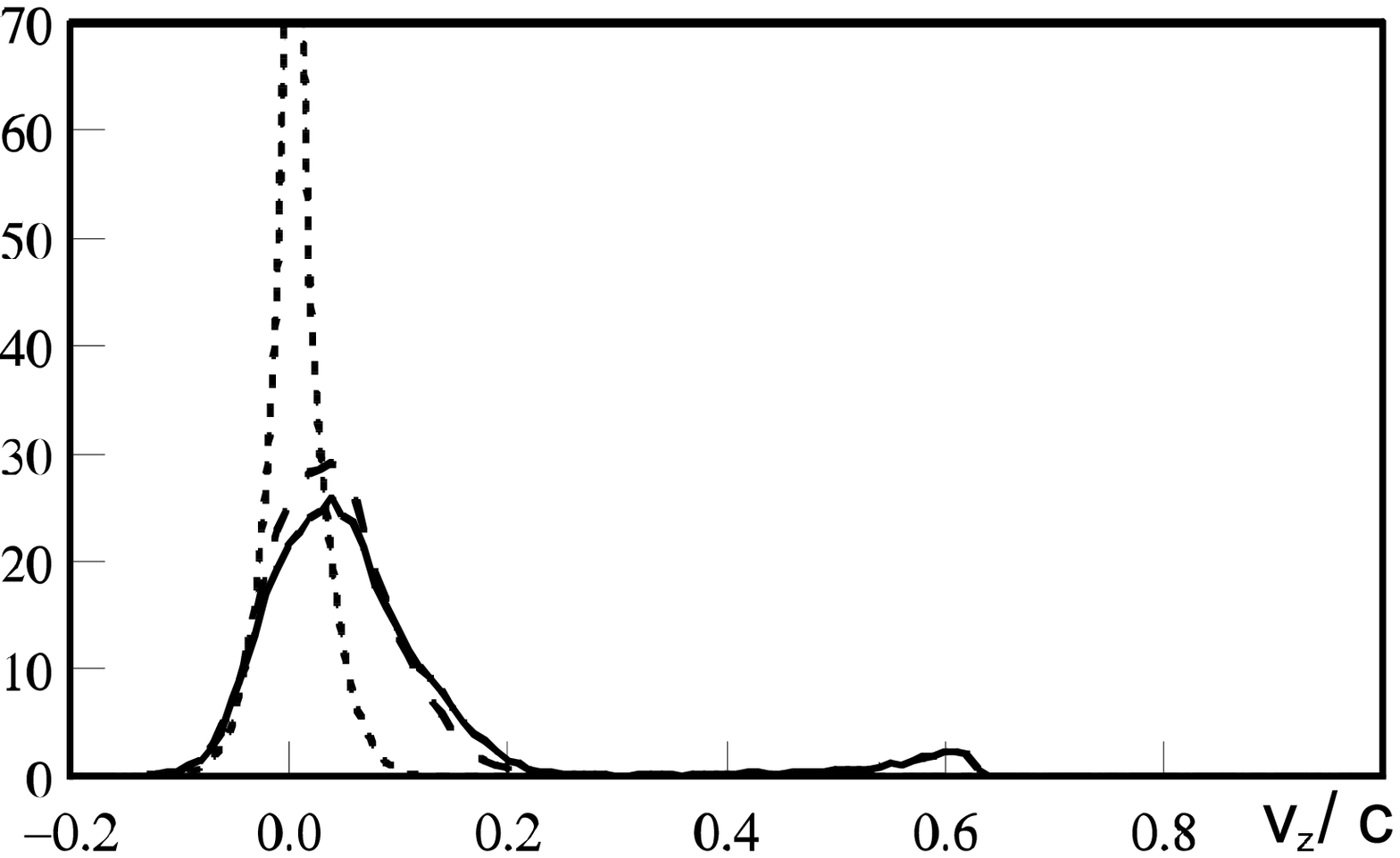}
\end{minipage}
\begin{minipage}{\columnwidth}
\includegraphics[width=0.65\columnwidth]{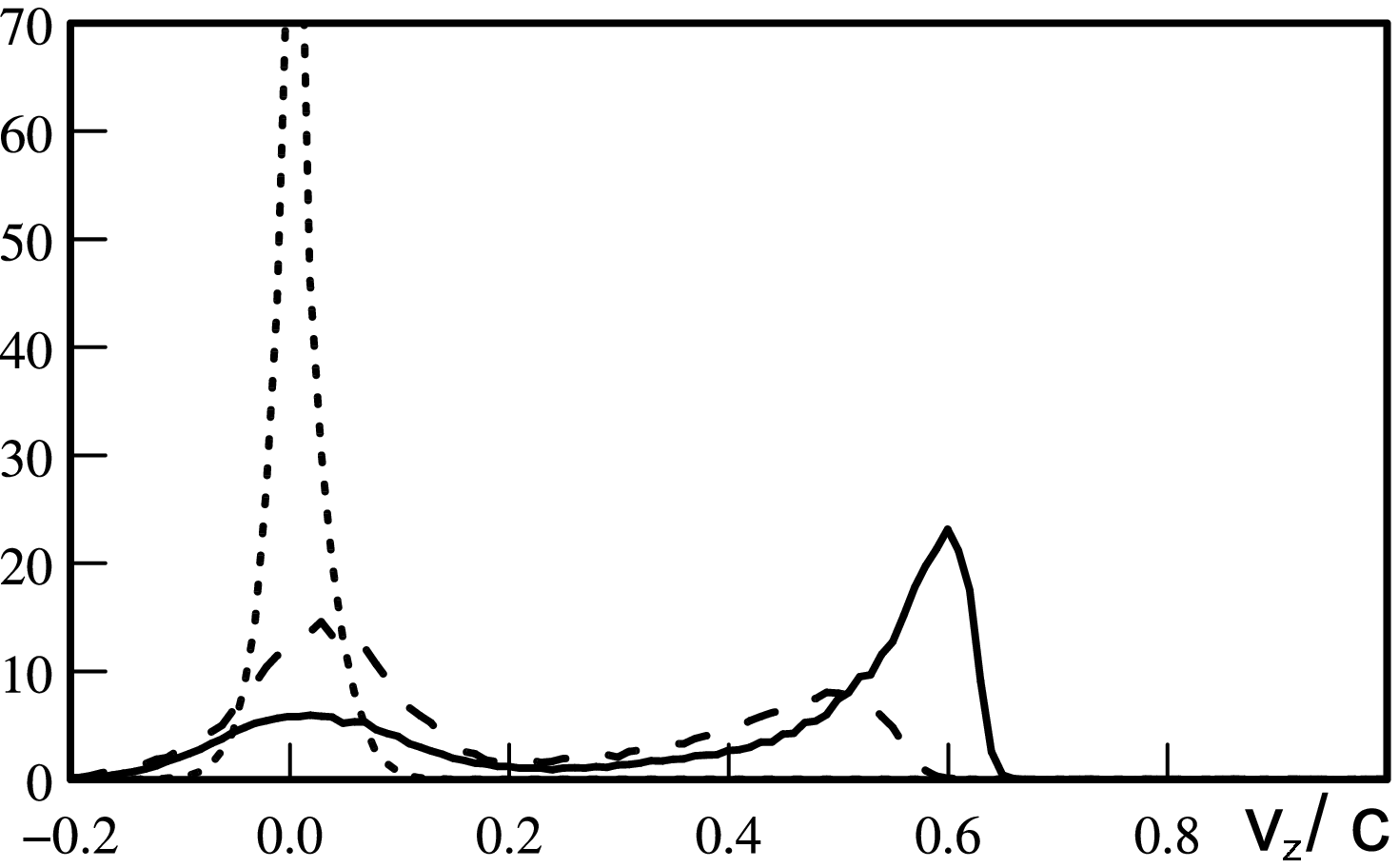}
\end{minipage}
\caption{Final probability distribution of the
parallel velocity for $\theta=15^\circ$ and $\overline{A}_1=0.1$
for the one-wave scheme with  $\overline{A}_2=0$ (top) and the
two-wave-scheme with $\overline{A}_2=0.01$ (bottom). The solid
lines ($e_0=0.3$), dashed lines  ($e_0=0.6$) and dotted lines
($e_0=1.99$) correspond to increasing densities.} \label{pdf}
\end{figure}

The Hamiltonian equations for the $\bf r$ and $\bf p$
are solved using a 4th order Runge-Kutta method. The time step
is adapted to ensure that the solution of a redundant evolution
equation for $H$ remains close to the expression~(\ref{Ham}).
The initial velocity
distribution has a temperature of the order of 1 keV. A population
of $5\ 10^{4}$ electrons is used in each simulation. Although
running the code with larger populations is not an issue, no
further information is derived from these larger runs, except of
course more converged statistics.

Numerical results confirm that the angle dependency
is not trivial and that parallel propagation ($\theta_1=0^\circ$)
for the primary wave is not always optimal~\cite{medrpa87}. Three
sets of simulations are presented hereafter. The parameters for
these simulations are relevant in today tokamak plasma. In
particular, the primary wave corresponds to the second harmonic of
the cyclotron frequency and the secondary wave to the third
harmonic Right Hand Polarized modes which are frequently used for
instance in the TCV tokamak experiments~\cite{algohe02}. The value
of the constant magnetic field is 1.42 $T$ for all simulations.
Three electron densities have been considered: $n_e \approx 0.6\
10^{19}\ m^{-3}$ ($e_0=0.3$), $n_e \approx 1.2\ 10^{19}\ m^{-3}$
($e_0=0.6$) and $n_e \approx 3.9\ 10^{19}\ m^{-3}$ ($e_0=1.99$). The
wave amplitudes are $\overline{A}_1=0.1$ and
$\overline{A}_2=0.01$ in the two-wave scheme and
$\overline{A}_2=0$ in the one-wave scheme.  They
correspond to power fluxes which are by orders of
magnitude higher than that achievable on gyrotron used in today
tokamak. However, preliminary computations using the Resonant Moment Method
suggest three-dimensional electromagnetic wave configurations
are very promising for larger acceleration of charged particles in
an external magnetic field with even lower wave amplitudes. In
such a case the wave vectors and the magnetic field are not
supposed to be co-planar and create a fully three dimensional
system. On the other side,  the required powers might be
achievable by free electron maser~\cite{thu02} even for
experiments with 2D launch configuration as predicted by our
direct particle simulations.

Also, the parameters $e_0=(\omega_{pe}/\Omega)^2=0.1-0.3$, correspond to the nighttime
ionosphere at approximately 130km, A=0.1 - to a power flux $5
W/cm^2$, and a frequency 2.6MHz of the primary wave. These
parameters are close to ones considered in \cite{medrpa87,
medrpa88} for single wave acceleration.

Figures~\ref{current-theta} show a significant increase of the
average parallel velocity for $e_0=0.3$ due to the secondary wave.
Moreover, the angle dependency of the average parallel velocity
appears to be maximal in the range $\theta_1=10^\circ - 60^\circ
$. This range thus appears to significantly differ from the
privileged value of the RM $I^*$ (Fig.2). This is not too
surprising since the RM have been computed assuming a Maxwellian
distribution with zero mean. Thus, although the global picture
from the RM description is reasonable, a more precise iterative
approach, taking into account the averaged velocity suggested by
the RM would be required for more accurate predictions. The
probability distributions of parallel velocity (Figures~\ref{pdf})
observed at the end of the simulations $\Omega\ t=2\ 10^4$
indicate that, in the two-wave scheme a much larger number of
electrons have had the occasion to interact with the primary wave
and the distributions of velocity exhibit two well marked maxima.
For $e_0=1.99$, the density is very close to the cut-off value of
the wave propagation and almost no effect is observed in both the
one-wave and the two-wave scheme. Also, if $\overline{A}_1$ is too
small, no averaged parallel velocity is observed at all.
Preliminary tests seem to reproduce the threshold previously
observed~\cite{kaakom90,cocomc91}.

This paper presents a mechanism for enhancing acceleration of a
population of electrons using crossing electromagnetic waves
propagating at different angles with respect to an external
magnetic field in a dispersive medium. The existence of optimal
angles of propagation for the primary wave is suggested using the
evaluation of resonant moments and is confirmed by direct
numerical simulations of the electron trajectories.  A secondary
low amplitude perpendicular wave is used to induce a
stochastization of the electron trajectories and, consequently, to
maintain a pseudo-equilibrium. Although measures of the
distributions (Figures~\ref{pdf}) clearly show a departure from a
thermal equilibrium, the stochastization effect of the secondary
wave yields a clear increase of the average parallel electron
velocity. It is a quite promising result since the amplitude of
the secondary wave is ten times lower the one of the first wave.

The authors are grateful to  Professor R. Balescu, Dr. B. Weyssow,
Dr. I Pavlenko, and Dr. R. Kamendje for useful discussions. D.C.
is researcher of the Fonds National pour la Recherche Scientifique
(Belgium). This work has been supported by the contract of
association EURATOM - Belgian state. The content of the
publication is the sole responsibility of the authors and it does
not necessarily represent the views of the Commission or its
services.

\end{document}